\documentclass{article}

\usepackage{arxiv}
\pdfoutput=1
\usepackage[utf8]{inputenc} 
\usepackage[T1]{fontenc}    
\usepackage[usenames,dvipsnames]{color}
\usepackage{hyperref}   
\hypersetup{
	colorlinks=true,
	linkcolor=black,
	citecolor=black,
	filecolor=black,
	urlcolor=black,
}

\usepackage{url}            
\usepackage{booktabs}       
\usepackage{amsfonts}       
\usepackage{nicefrac}       
\usepackage{microtype}      
\usepackage{lipsum}		
\usepackage{graphicx}
\usepackage[round]{natbib}
\usepackage{doi}
\usepackage{multirow}
\usepackage{enumitem}
\usepackage{graphics}
\usepackage{float}
\usepackage{siunitx}
\usepackage{xspace}
\usepackage{amsmath}
\usepackage{tabularx}
\newcolumntype{s}{>{\hsize=.5\hsize}X}

\def\dodoi#1{doi: \href{https://doi.org/#1}{\nolinkurl{#1}}}

\title{Lower Interaural Coherence in Off-Signal Bands Impairs Binaural Detection}

\author{\href{https://orcid.org/0000-0002-9477-7106 }{\includegraphics[scale=0.06]{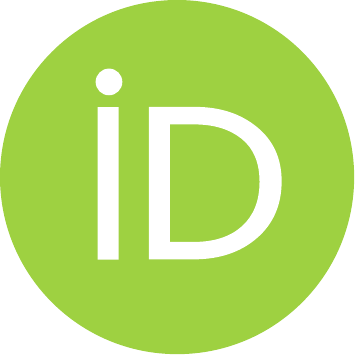}\hspace{1mm}Bernhard Eurich} \\
	Department für Medizinische Physik und Akustik\\
	Universität Oldenburg\\
	26111 Oldenburg, Germany \\
	\texttt{bernhard.eurich@uni-oldenburg.de} \\
	\And
	\href{https://orcid.org/0000-0002-6067-1602 }{\includegraphics[scale=0.06]{orcid.pdf}\hspace{1mm}Jörg Encke}\\
	Department für Medizinische Physik und Akustik\\
	Universität Oldenburg\\
	26111 Oldenburg, Germany \\
	
	\And
	\href{https://orcid.org/0000-0002-1644-4947  }{\includegraphics[scale=0.06]{orcid.pdf}\hspace{1mm}Stephan D. Ewert} \\
	Department für Medizinische Physik und Akustik\\
	Universität Oldenburg\\
	26111 Oldenburg, Germany \\
	
	\And
	\href{https://orcid.org/0000-0002-1830-469X  }{\includegraphics[scale=0.06]{orcid.pdf}\hspace{1mm}Mathias Dietz}\\
	Department für Medizinische Physik und Akustik\\
	Universität Oldenburg\\
	26111 Oldenburg, Germany \\
	
}

\renewcommand{\headeright}{}
\renewcommand{\shorttitle}{Lower Interaural Coherence in Off-Signal Bands Impairs Binaural Detection -- Preprint}

\hypersetup{
	pdftitle={Interaural Coherence Across Frequency Channels Accounts for Binaural Detection in Complex Maskers},
	pdfauthor={Bernhard Eurich, Jörg Encke, Stephan D. Ewert, Mathias Dietz},
}

\begin{document}
	\maketitle
	
	\begin{abstract}
		Differences in interaural phase configuration between a target and a masker can lead to substantial binaural unmasking. This effect is decreased for masking noises with an interaural time difference (ITD). Adding a second noise with an opposing ITD in most cases further reduces binaural unmasking. Thus far, modeling of these detection thresholds required both a mechanism for internal ITD compensation and an increased binaural bandwidth. An alternative explanation for the reduction is that unmasking is impaired by the lower interaural coherence in off-frequency regions caused by the second masker (\citealp{marquardt2009}, JASA pp. EL177 - EL182). Based on this hypothesis, the current work proposes a quantitative multi-channel model using monaurally derived peripheral filter bandwidths and an across-channel incoherence interference mechanism. This mechanism differs from wider filters since it has no effect when the masker coherence is constant across frequency bands. Combined with a monaural energy discrimination pathway, the model predicts the differences between a single delayed noise and two opposingly delayed noises, as well as four other data sets. It helps resolve the inconsistency explaining some data sets requires wide filters while others require narrow filters.    
	\end{abstract}

	
\section{\label{intro}Introduction}
The detection of a pure tone in noise is facilitated by differences in the interaural phase between tone and noise \citep{hirsh1948}. The improvement in the detection threshold compared to the diotic case is referred to as the binaural masking level difference (BMLD). The maximum BMLD is observed when detecting an antiphasic pure tone target ($S_{\pi}$) in an in-phase noise masker ($N_0$). Adding an interaural time difference (ITD) to the masker has been observed to reduce the BMLD \citep{langford1964}. A particularly simple case is when the noise and the target tone have exactly opposite interaural phase differences. In this case, detection thresholds increase gradually and monotonically with increasing noise ITD \citep{rabiner1966}. The increase can be simulated accurately by exploiting changes in the cross-correlation coefficient of left and right signal after using a filter with an equivalent rectangular bandwidth (ERB) of 60\ldots\,85\,Hz at a center frequency of 500\,Hz \citep{rabiner1966, dietz2021}. This bandwidth range resembles the established estimate of the human peripheral filter bandwidth obtained from monaural psychoacoustic experiments at this frequency which is 79\,Hz \citep{glasberg1990} and referred to as standard filter bandwidth in the following.
\begin{figure}[ht]
	\includegraphics[width=\columnwidth]{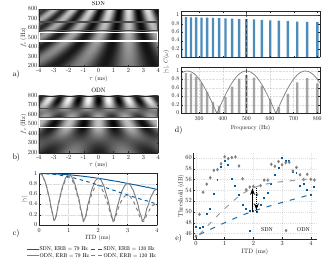}
	\label{figExplain}
	\caption{ \textbf{a)} Cross-correlogram of delayed noise (SDN) with ITD = 2\,ms. White and black areas represent maxima and minima of the cross-correlation functions, respectively. The white box highlights the 500\,Hz frequency channel while the gray box highlights a channel centered at 625\,Hz. \textbf{b)} Interaural cross-correlogram as in (a) but for opposingly delayed noises (ODN). \textbf{c)} Interaural coherence $|\gamma|$ as a function of noise ITD for SDN (blue lines) and ODN (gray lines) for two underlying filter bandwidhts. \textbf{d)} Continuous lines: Normalized cross-power spectral density (CPSD) at ITD~=~2\,ms as a function of frequency, $C(\omega)$, as derived in Eq.\,\ref{eqcos1} et seq.; Bars: Interaural coherence $|\gamma|$ of the signals after peripheral Gammatone filtering. \textbf{e)} Thresholds of $S_{\pi}$ detection in SDN and ODN as a function of ITD from \citet{vanderheijden1999}. The dashed lines symbolize the coherence-decline-induced threshold increase determined by a binaural bandwidth of ERB = 79\,Hz (lower line) and ERB = 130\,Hz (upper line). As denoted by the arrows, the data can be explained in two ways: (1: dotted downward arrow) The ODN thresholds are determined by the cross-correlation function at 500 Hz and a binaural bandwidth $\geq$ 130\,Hz. A delay line causes the lower SDN thresholds. (2: solid upward arrow) The SDN thresholds are determined by the ITD-dependent coherence as derived from an ERB of 79\,Hz. Off-frequency incoherence in ODN causes higher ODN thresholds.}
\end{figure}

Another explanation uses an array of different internal delays, known as delay lines \citep{vanderheijden1999, stern1978, bernstein2018, bernstein2020c}. \citet{jeffress1948} suggested that the binaural system has the ability to compensate for the external ITD. The compensation accuracy or efficiency has been assumed to decrease with masker ITD in order to model the decreasing BMLD \citep{stern1978, vanderheijden1999, bernstein2017a}.

\citet{vanderheijden1999} generated a new stimulus which they termed double-delayed noise (diamonds in Fig.\,\ref{figExplain}e) by adding two noises, one with a positive and one with a negative ITD. We refer to this as opposingly delayed noises (ODN). They found detection thresholds in ODN to be substantially higher than in "regular" delayed noise termed "single-delayed noise", SDN. Since internal delays can only compensate for the ITD of one noise, ODN limits the usefulness of the putative delay lines. Thus, \citet{vanderheijden1999} attributed the additional unmasking in SDN, compared to ODN, to the delay lines. So far, only models based on delay lines have precisely accounted for both SDN and ODN detection thresholds. The SDN-ODN detection threshold difference is therefore used as psychoacoustic evidence for long delay lines \citep{stern2019}. The difference was in fact used to derive the length and potency of the delay line system \citep{vanderheijden1999}.

However, two problems exist with establishing the psychoacoustically derived delay line length or internal delay distribution function. First, measured delays in binaural neurons of mammals are short compared to the respective period duration  \citep{mcalpine2001, joris2006} and thus too short to fulfill the lengths requirements of delay line models \citep{thompson2006, marquardt2009, stern2019}. 

Second, if the delay-line models use their internal delays to account for SDN thresholds while correlation coefficient-based models \citep{rabiner1966, dietz2021,encke2021} are equally precise for SDN without delay lines, the two model types must differ by something else: filter bandwidth. \citet{vanderheijden1999} started off with ODN to determine the filter bandwidth. They could best fit their ODN thresholds with filters of various shapes and an ERB of 130 to 180\,Hz at 500\,Hz center frequency. This is expectedly larger than what models without delay lines, such as \citet{encke2021}, required for SDN. The two versions cannot both be correct. Thus, either the SDN-threshold-based filter bandwidth is confounded by not considering delay lines or the ODN-based filter bandwidth fit by \citet{vanderheijden1999} is confounded by something else. For the latter, \citet{marquardt2009} offered a possible explanation. They identified the interaural coherence to be lower in certain off-frequency regions in ODN but not in SDN. They argued that the higher detection thresholds in ODN could also originate from some detrimental off-frequency impact related to the low coherence rather than from a wider filter bandwidth per se (upward arrow in Fig.\,\ref{figExplain}e). If this is true, both SDN and ODN thresholds can potentially be predicted using the same standard filter bandwidth.  Fig.\,\ref{figExplain}, panels a), b) and d), show that the cross-power spectral density is constant across frequency in SDN but spectrally modulated in ODN (see Appendix for derivation).

Leaving aside the first physiologic argument, there are two options to account for the SDN-ODN difference. (1) wider filters combined with delay lines (downward arrow in Fig.\,\ref{figExplain}e) or (2) filters with standard peripheral bandwidths and a detrimental off-frequency impact (upward arrow in Fig.\,\ref{figExplain}e). However, recent data of SDN thresholds measured for different noise bandwidths can only be accurately simulated with filters falling into the standard peripheral bandwidth category \citep{bernstein2020c,dietz2021}, causing a logical impasse for the wider-filters assumption even within the psychoacoustic domain and for SDN alone.

The aim of this study is thus to develop a model that accounts for SDN and ODN thresholds at the same time, using a standard filter bandwidth and -- consequently -- without long delay lines but with off-frequency impact.	Specifically, we suggest an across-frequency incoherence interference mechanism which is inspired by binaural interference \citep{bernstein1995} and modulation detection interference \citep{yost1989, oxenham2001}. This makes sure that the same "hardware" causes different detection thresholds for maskers with different amounts of off-frequency IPD fluctuations. The here developed mechanism will be described in Section \ref{descr} and used to predict critical binaural detection data in Section \ref{pred}.

Even beside the discussion concerning delay lines in humans and other mammals, the width of filters has caused an unresolved contradiction in the binaural literature that filters need to be narrow to account for some and broad to account for other data (see \citealp{verhey2020} for a review). Generally speaking, detection thresholds in spectrally simpler maskers can be simulated using a standard peripheral filter bandwidth \citep{breebaart2001b}, whereas more complex maskers appear to be processed by wider filters or alternative across-frequency processes \citep{kolarik2010}. We therefore evaluated our model with data from five different studies in three groups: 
\begin{enumerate}
	\item \citet{vanderheijden1999} combined all key aspects required to revisit \citeauthor{marquardt2009}'s hypothesis: (a) The SDN thresholds are planned to be determined by the decay of $|\gamma|$ with a 79\,Hz-wide Gammatone filter. (b) The ODN thresholds supposedly will, despite the same 79\,Hz on-frequency filter, be elevated by the across-channel incoherence interference. The most important datapoints to judge this are at those ITDs where the on-frequency coherence of SDN and ODN is the same, but thresholds differ. For $S_0$ detection this is the case at ITD =~1\,ms and 3\,ms, for $S_{\pi}$ detection at ITD = 2\,ms and 4\,ms.
	\item \citet{marquardt2009} not only presented the above-mentioned hypothesis but also detection thresholds. There, SDN and ODN maskers are spectrally surrounded by bands that each have a different, constant IPD. Due to the different signs they either do or do not cause interaural incoherence at the transitions. Their reported differences impose a challenge for single-channel models that use a constant filter bandwidth. 
	\item \citet{sondhi1966}, \citet{holube1998} and \citet{kolarik2010} reported detection thresholds of an $S_{\pi}$ tone centered in an in-phase noise that is spectrally surrounded by antiphasic noise. These simulations are included for an additional discussion about the proposed standard-filter-plus-off-frequency-impact concept, since larger binaural bandwidths have previously been derived based on such data.
\end{enumerate}

\section{\label{descr}Description of the Model}

\begin{figure}[ht]
	\centering
	\includegraphics[width=0.5\columnwidth]{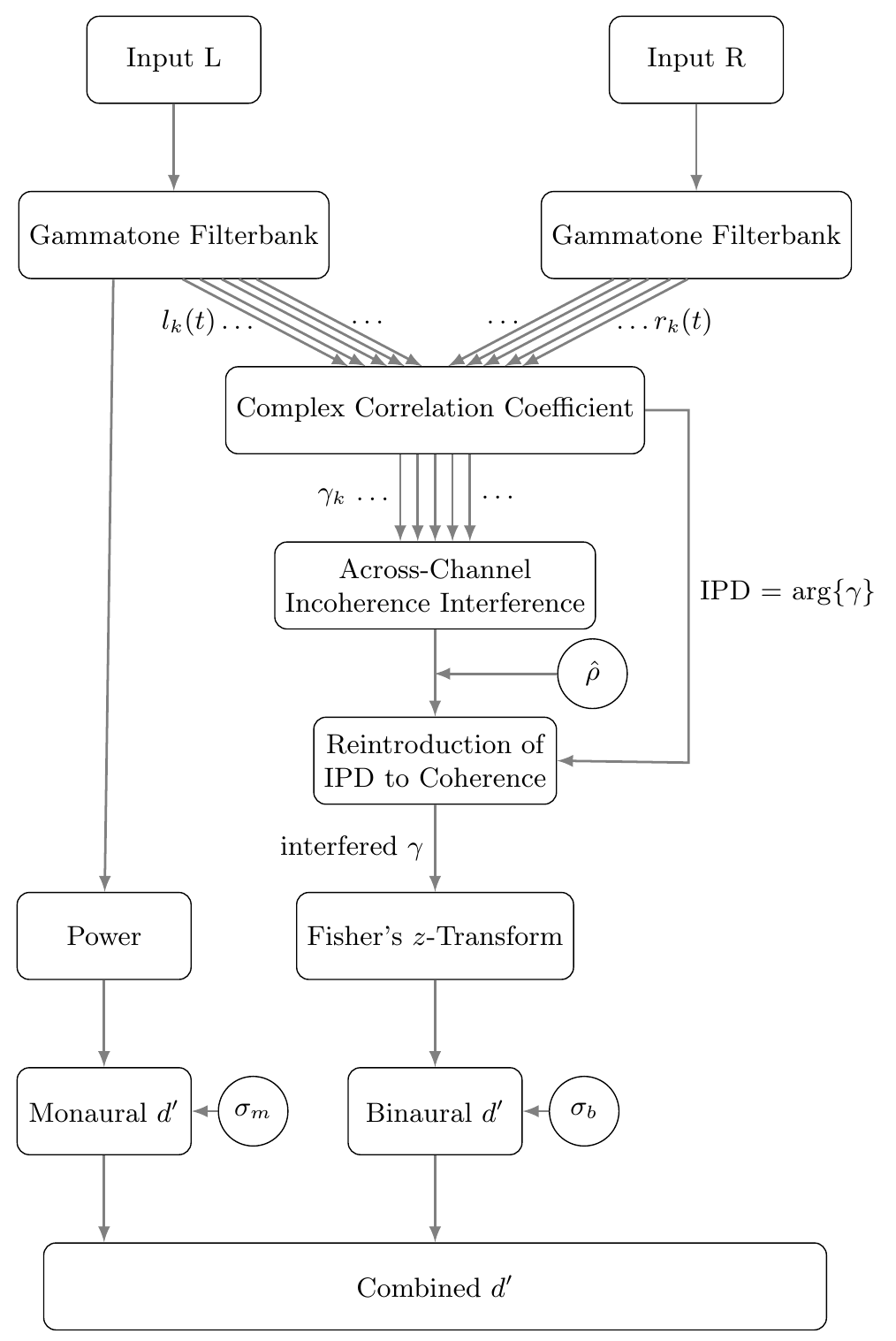}
	\caption{\label{figFlowchart}{Processing stages of the proposed model. See main text for details.}}
\end{figure}

Fig.\,\ref{figFlowchart} shows the processing stages of the proposed model. It is designed as a numerical multi-channel model through all stages, but these were here realized and tailored to predict binaural-detection data with a 500\,Hz pure-tone target. The model builds on the analytical single-channel model approach of \citet{encke2021}. It furthermore includes an across-frequency incoherence interference mechanism. It consists of a multi-channel binaural processing pathway and a monaural pathway. Both pathways compare multiple tokens of the processed representation of the condition-specific masker only to the representation of signal plus masker. This comparison has been suggested to mimic a subject's strategy of comparing a stimulus to a learned reference template \citep{dau1996, dau1997b, jepsen2008, breebaart2001, bernstein2017a}. Based on these comparisons, both pathways deliver a sensitivity index ($d'$). An optimal combination of the pathways' estimates gives the overall $d'$ estimate of the model \citep{green1966a, biberger2016}.

\subsection{\label{periph}Peripheral Processing}
The left and right input signals were processed with a fourth-order Gammatone filterbank that represents basilar-membrane bandpass filtering. The filterbank implementation by \citet{hohmann2002} was employed with a spacing of five filters per ERB in the range of 67\,Hz to 1000\,Hz. The grid was defined by centering one filter at 500\,Hz. This filter had an ERB of 79\,Hz \citep{glasberg1990} and was indexed with $k=0$.

To focus on the impact of the spectral masker properties discussed above, the present implementation did not include any other peripheral processing such as low-pass filtering, power-law compression or half-wave rectification. Only Gaussian noise was used as masker, and only 500-Hz tones as targets. 

\subsection{\label{binpath}Binaural Pathway}
The correlation coefficient $\gamma(\tau=0) = \gamma$ was derived from the analytical, i.e. complex-valued left and right signals  $l(t)$ and $r(t)$ in the frequency channel $k$, provided by the Gammatone filterbank:
\begin{equation}
	\gamma_k = \frac{\overline{l_k(t)^* \,r_k(t)}}{\sqrt{ \overline{|l_k(t)|^2} \overline{|r_k(t)|^2}}}
\end{equation}
where $\overline{\bullet}$ marks the temporal mean. This results in one complex correlation coefficient per frequency channel, averaged over the whole stimulus duration. The complex-valued correlation coefficient was used because it conveniently combines information about both the mean IPD as $\arg\{\gamma\}$ and about the amount of IPD fluctuations in the form of interaural coherence $|\gamma|$. While the Introduction mentioned a mismatch between mammalian physiology and delay line models, it should be noted that the seemingly abstract use of complex-valued correlation is identical to two real-valued correlations with a 90° phase offset. Such two orthogonal correlators exist in the form of the average left- and right hemispheric binaural neuron in mammals \citep{mcalpine2001, joris2006}. The physiologic relation of $\gamma$ is explained in more detail in \citet{encke2021}. 

As pointed out in the Introduction, the novelty of the present model is the interference of IPD fluctuations across frequency channels. The term incoherence interference describes purely detrimental effects, meaning only channels with lower coherence affect their neighborhood, but not the other way around. This process is implemented as a \emph{restricted} across-channel weighted average of the coherence $|\gamma|_{k}$: The $|\gamma|_k$ are limited such that they can no more exceed the on-frequency $|\gamma|$, thus referred to as $|\gamma|_{k, \textrm{lim}}$. A weighted average of $|\gamma|_{k, \textrm{lim}}$ therefore means that only channels with lower coherence, or stronger IPD fluctuations, affect the on-frequency coherence, i.e. interference of IPD fluctuations:
\begin{equation}
	|\gamma|_w = \sum_{-m}^{m} w(k) |\gamma|_{k, \textrm{lim}}
\end{equation}
$w(k)$ symbolizes a function that weights the contribution of a channel $k$ to the resulting $|\gamma|_w$. The employed weighting function has an exponential decay described by
\begin{equation}
	w(k) = e^{-|k|/(b\,\sigma_w)}.
\end{equation}
$\sigma_w$ represents the decay parameter, normalized by the number of filters per ERB, $b$. The double-exponential decay shape was chosen by empirical trials. While the exact shape of the window was not crucial, we did not obtain more precise simulations with other shapes.

For a low masker coherence, or at the practically irrelevant case of a positive SNR, adding a target with an IPD of $\pi$ relative to the masker can swap the mean IPD from the masker to that of the target. In special cases, the masker alone and masker plus target can have the same coherence, but differ in their mean IPD and thus in their correlation. Thus, the interaural coherence $|\gamma|$ is not sufficient as a decision variable. Instead, $\gamma$, including both coherence and the mean IPD, is required. Therefore, the original mean IPD is now reintroduced to the coherence after the limitation and interference stage, so that the model can operate on the \emph{complex} correlation coefficient as suggested by \citet{encke2021}. 

\begin{equation}\label{eqgammaw}
	\gamma_w = |\gamma|_w \,e^{\arg\{\gamma_0\}}
\end{equation}

Unity-limited measures such as coherence or correlation can be Fisher $z$, i.e. $\textrm{atanh}$-transformed for the purpose of variance normalization (\citealp{mcnemar1969, just1994}, as often applied in psychophysics, e.g., \citealp{luddemann2007, bernstein2017a}). As in \citet{encke2021}, $\gamma_w$ is multiplied by a model parameter $\hat{\rho} < 1$ to avoid an infinite sensitivity to deviations from a coherence of one. This is equivalent to adding uncorrelated noise to the two input signals. The decision variable of the binaural pathway is thus
\begin{equation}
	\zeta = z[\hat{\rho} \,\gamma_w] 
\end{equation}
where $z[\bullet]$ is the Fisher $z$-transform applied to the modulus of $\gamma_w$ while leaving the argument unchanged. 

In the signal detection stage, the $d'$ is obtained based on the difference between the ensemble averages of the representations of the target signal plus noise, $\zeta_{N+S}$, and the representations of the noise alone, $\zeta_N$:
\begin{equation}
	d'_b = \frac{|\zeta_{N+S} - \zeta_N|}{\sigma_b}
\end{equation}
The internal noise $\sigma_b$ defines the sensitivity of the binaural model pathway \citep{dietz2021}.

\subsection{\label{monpath}Monaural Pathway}
For the monaural pathway, the power $P$ of the on-frequency filter channel was evaluated. It is half the squared mean of the envelope across the whole signal duration \citep{biberger2016}. The envelope is the modulus of the complex-valued filter output:
\begin{equation}
	P = \frac{\overline{|u_0(t)}|^2}{2}.
\end{equation}
In the stimuli employed in this study, the power is identical in the left and right channels, thus it is sufficient to evaluate only one side.

For a signal-induced power change $\Delta P = P_{N+S} - P_{N}$, the processing accuracy is limited by a level-dependent internal noise with a Gaussian distribution of amplitudes and a standard deviation of $\sigma_m$. Thus, the sensitivity of the monaural pathway is equivalent to	
\begin{equation}
	d'_m = \frac{\Delta P / P_{\textrm{avg}}}{\sigma_m},
\end{equation}
where $ P_{\textrm{avg}}$ represents the average power between $P_{N+S}$ and $P_{N}$.

\subsection{\label{det}Detector}
The sensitivity indices of the binaural, $d'_b$, and monaural pathway, $d'_m$ were combined assuming two independent information channels \citep{green1966a,biberger2016}
\begin{equation}
	d'_{b+m} = \sqrt{d'^2_b + d'^2_m}.
\end{equation}
The $d'$ that corresponds to the experiment-specific detection thresholds was obtained via table-lookup (Numerical evaluation in \citealp{hacker1979}). This depends on the number of intervals, as well as the specific staircase procedure used in the simulated experiments. For each condition, the model was evaluated for a range of target levels. This delivered the psychometric function. The predicted detection threshold was obtained from a straight line fitted to the logarithmic $d'$.
The model parameters were manually adjusted in order to optimize the prediction accuracy. The resulting parameter values are given in Table\,\ref{tabexp}.

\section{\label{pred}Predictions of Binaural-Detection Datasets}
In all experiments, a 500\,Hz $S_{\pi}$ or $S_0$ tone was to be detected in a broadband Gaussian noise masker. Figures \ref{figvdH}, \ref{figMc} and \ref{figKol} show the experimental data denoted by symbols, the predictions of the proposed model including incoherence interference as continuous lines, as well as predictions of the single-channel version which means without incoherence interference, as dotted lines.
Three types of binaural-detection experiments were simulated, as described in detail in the following subsections.	
Table \ref{tabexp} summarizes the parameter values used to simulate the experimental conditions. It further lists the non-adjusted coefficient of determination ($R^2$, interpretable as the proportion of variance in the data explained by the model) and the root-mean-square error (RMSE) of the simulations both with and without the proposed incoherence interference.
\begin{table*}[ht]
	
	\centering
	\resizebox{\textwidth}{!}{
		\begin{tabular}{ccccccc@{\hspace{.5cm}}cccc}
			\toprule
			\hline
			\multirow{2}{*}{Experiment}&\multirow{2}{*}{Signal}&\multirow{2}{*}{Variable}&\multirow{2}{*}{$\hat{\rho}$}&\multirow{2}{*}{$\sigma_b$}&\multirow{2}{*}{$\sigma_w$}&\multirow{2}{*}{$\sigma_m$}&\multicolumn{2}{c}{with}&\multicolumn{2}{c}{without}\\
			&&&&&&&$R^2$~&RMSE / dB &$R^2$& RMSE / dB\\
			\midrule
			\multirow{2}{*}{\citet{vanderheijden1999}}&$\pi$&\multirow{2}{*}{ITD}&0.91&0.20&0.50&0.40&0.94&0.85&0.78&1.45 \\
			&$0$&&0.86&0.17&0.65&0.40&0.87&0.86&0.57&1.38 \\
			\hline
			\citet{marquardt2009}&$0$&BW&0.89&0.24&0.65&0.40&0.96&0.37&-0.62&1.97\\
			
			\hline
			\citet{kolarik2010}&$0$&BW&0.91&0.20&0.50&0.40&0.97&0.67&0.42&3.07\\
			\hline
			\bottomrule
			
		\end{tabular}
	}
	\caption{\label{tabexp}Summary of the simulated experiments and predictions. \emph{Columns 1 - 3}: Simulated experiment, IPD of the used target signal, independent variable. \emph{Columns 4 - 7}: Used model parameters: $\hat{\rho}<1$: Maximum coherence (internal noise); $\sigma_b$: Standard deviation of the internal noise to determine the absolute performance of the binaural pathway; $\sigma_w$: Slope parameter of the double-exponential across-channel interaction window (normalized by the number of filters per ERB); $\sigma_m$: Standard deviation of the level-dependent internal noise to determine the accuity of the monaural pathway; \emph{Columns 8 - 11}: Accuracy of the predictions with and without incoherence interference: Coefficient of determination ($R^2$, interpretable as the proportion of the variance in the data explained by the model); root-mean-square errors (RMSE) of the predictions.}
\end{table*}

\subsection{\label{vdH}van der Heijden \& Trahiotis 1999}
\begin{figure}[ht]
	\centering
	\includegraphics[width=0.5\columnwidth]{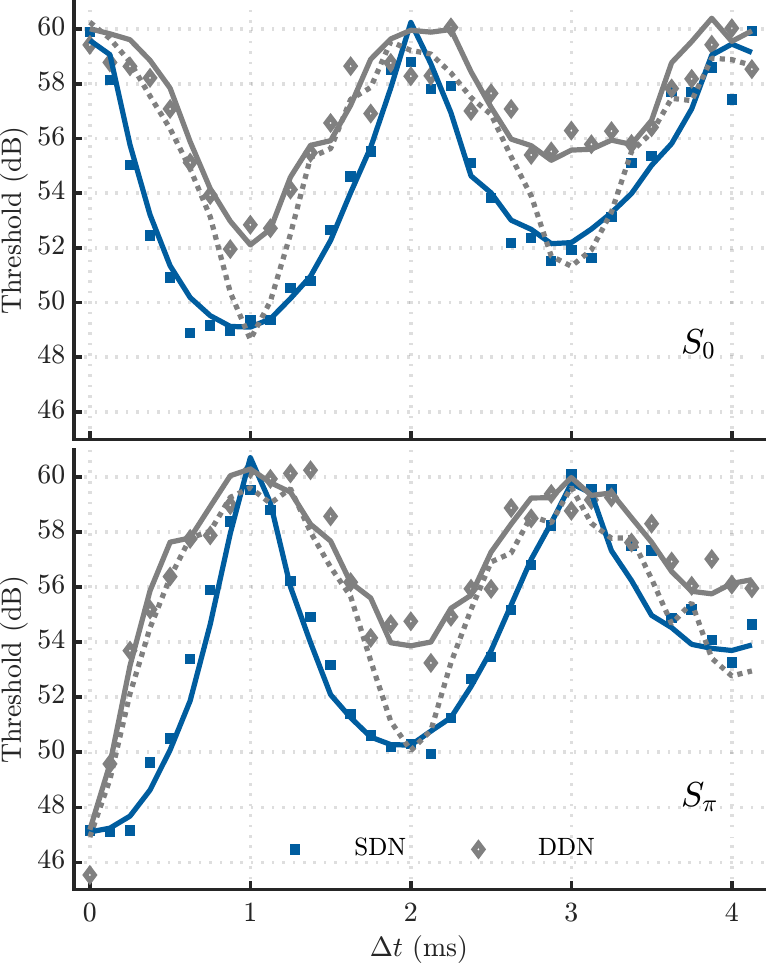}
	\caption{\label{figvdH}Experimental data from \citet{vanderheijden1999} (symbols). The continuous lines show the predictions of the presented model including the across-channel incoherence interference. The dashed lines show predictions for ODN without interference (single-channel version), equivalent to \citet{encke2021}. \textit{Upper panel:} Detection thresholds with $S_0$ target; \textit{lower panel:} $S_{\pi}$ target.}
\end{figure}

In this arguably most central experiment, detection thresholds of an $S_0$ target tone (Fig.\,\ref{figvdH}, upper panel) as well as of an $S_{\pi}$ tone (Fig.\,\ref{figvdH}, lower panel) were measured as a function of the interaural masker ITD in steps of 0.125\,ms. The bandwidth of the masker was 900\,Hz. As outlined in the Introduction, the ODN consisted of two superimposed noises with opposite ITD. 
The experiment performed by \citet{vanderheijden1999} employed a four-interval, two-alternative forced choice task (4I-2AFC, first and fourth intervals always contained only the masker and served as queuing intervals). Their adaptive 2-down 1-up stair case procedure estimated the 70.7\,\% correct-response threshold. This is equivalent to a $d'$ of 0.78 at threshold. Thus, as described in section \ref{det} the model determined the threshold in the form of the signal level producing this $d'$.
The continuous lines in Fig.\,\ref{figvdH} show the simulations of the presented model, including the across-channel incoherence interference.
From visual inspection, the simulations captured all effects from the experimental thresholds and the critical threshold differences between SDN and ODN at all ITDs under both conditions. Specifically, the critical threshold differences of 3.5\,dB at ITD~=~1\,ms in the $S_{0}$ condition and 4\,dB ITD~=~2\,ms in the $S_{\pi}$ condition are precisely accounted for. This good correspondence is also reflected in the around 90\,\% explained variance under both conditions, and RMS errors of less than 1\,dB.
The dashed lines show simulations without the across-channel incoherence interference (single-channel model, cf. \citealp{encke2021}) but all other model parameters unchanged. This shows that a large amount of the threshold differences is already explained by differences in the on-frequency coherence. In much the same way as ODN coherence oscillates as a function of analysis frequency (Fig. \ref{figExplain}d), it also fluctuates as a function of the masker ITD  (Fig. \ref{figExplain}c). Particularly at ITD~=~0.5\,ms, ODN is incoherent in the 500-Hz band, whereas SDN is almost fully coherent. This, and not the across-frequency process, causes the difference in the simulated thresholds at this ITD. The across-frequency process only comes into play at those ITDs where the coherence at 500\,Hz (on-frequency) is nearly identical in SDN and ODN (upper panel: ITD~=~1\,ms and 3\,ms; lower panel: ITD~=~2\,ms and 4\,ms).
\subsection{\label{MqMc}Marquardt \& McAlpine 2009}

\begin{figure}[ht]
	\centering
	\includegraphics[width=0.5\columnwidth]{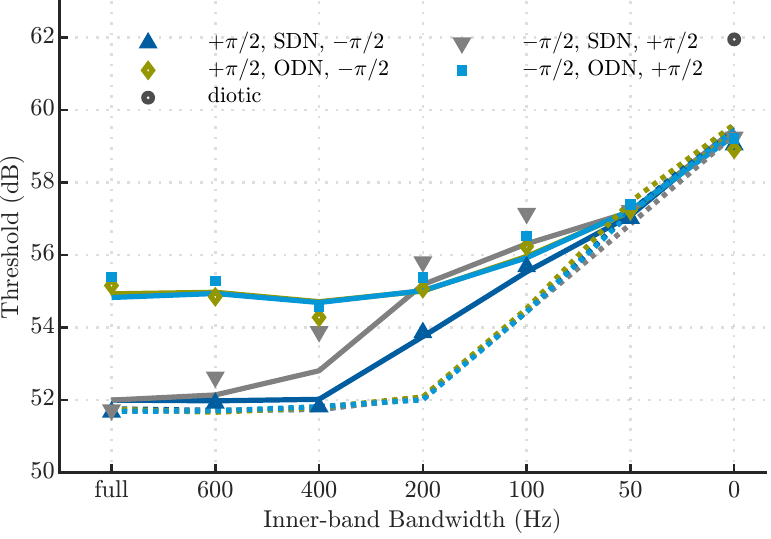}
	\caption{\label{figMc}Experimental data from \citet{marquardt2009} (symbols) and model predictions (lines). Detection thresholds are given as function of the inner-band bandwidth. The inner band contains delayed noise (triangles) or opposingly delayed noises (diamonds and bullets) with a fixed ITD = 1\,ms while the flanking bands have a constant phase shift of $+\pi/2$ (upward triangle and diamond) and $-\pi/2$, or vice versa (downward triangle and bullet). Continuous and dashed lines again show predictions with and without  across-frequency incoherence interference, respectively.}
\end{figure}

The masker of this experiment contained SDN and ODN centered at the frequency of the $S_0$ target tone with a constant ITD = 1\,ms in the inner band. The inner band was spectrally surrounded by bands that each had a constant IPD of $\pi/2$ and $-\pi/2$, or vice versa. Thresholds are given as a function of the inner-band bandwidth.
The resulting phase transitions between inner and flanking bands have been hypothesized to impair the detection if they cause a frequency region of low interaural coherence.
The lower and upper frequency limits of the composite stimuli are 50\,Hz and 950\,Hz, respectively. The two-interval-two-alternative-forced choice task with a 3-down 1-up procedure that was used estimated the thresholds to be 79.4\,\% correct. This corresponds to $d' = 1.14$ at the threshold predicted by the model.	
In Fig.\,\ref{figMc}, detection thresholds of the $S_0$ tone are shown as a function of the bandwidth of the inner band. Again, the model predicted all critical characteristics of the data. These are the 3\,dB difference between SDN and ODN at the full inner-band bandwidth (same as ITD~=~1\,ms in the $S_0$ condition in \citealp{vanderheijden1999}), the elevated SDN thresholds in the [-$\pi/2$, SDN, +$\pi/2$] compared to the [+$\pi/2$, SDN, -$\pi/2$] condition and the 3\,dB BMLD where the inner-band bandwidth is zero. Without the incoherence interference, the predictions cannot distinguish between the different conditions of the experiment. They deviate more from the mean than the data, resulting in a negative $R^2$.

\subsection{\label{Kol}Experiments on the operating bandwidth in binaural detection}
\begin{figure}[ht]
	\centering
	\includegraphics[width=0.5\columnwidth]{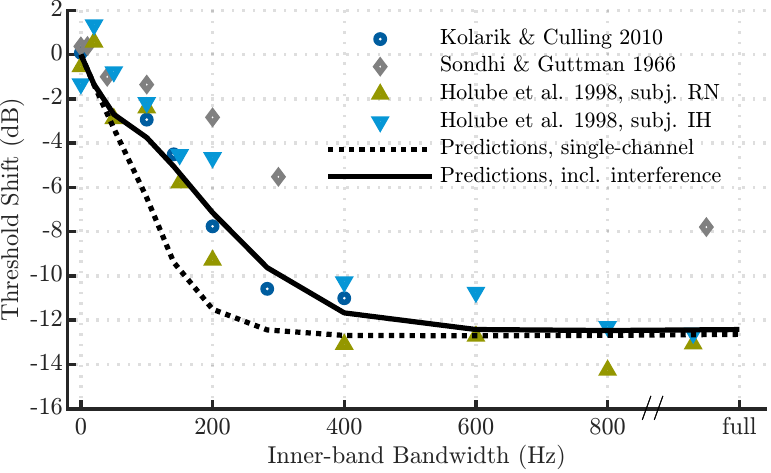}
	\caption{\label{figKol}{Symbols denote data from binaural detection experiments with the configuration $N_{\pi0\pi} S_{\pi}$ as a function of the inner-band ($N_0$) bandwidth; continuous and dotted line: Model prediction with and without across-incoherence incoherence interference.}}	
\end{figure}

Several studies investigated the operating bandwidth in binaural detection using maskers that contains two flanking bands which differ in their interaural configuration from the inner band \citep{sondhi1966, holube1998, kolarik2010}. 
The masking noise is diotic ($N_0$) in the inner band and antiphasic ($N_{\pi}$) in the flanking bands. Detection thresholds of an $S_{\pi}$ target tone were again measured as a function of the inner-band bandwidth. Results are expressed as the difference between thresholds in the flanked condition and the threshold without an inner band, i.e. $N_{\pi} S_{\pi}$. In Fig.\,\ref{figKol}, the circles mark the threshold differences reported by \citeauthor{kolarik2010} (\citeyear{kolarik2010}, centered condition), which represent averages across their three participants. The triangles show individual thresholds of the two participants in the study by \citeauthor{holube1998} (\citeyear{holube1998}, rectangular condition). The gray diamonds show the data from \citet{sondhi1966}. Our model predictions were oriented on the 2-down 1-up 2I-2AFC paradigm employed in \citet{kolarik2010}, equivalent to $d'=0.78$ at threshold. The black continuous line shows the model predictions with the same parameter settings as used to predict the $S_{\pi}$ detection thresholds in \citeauthor{vanderheijden1999} (\citeyear{vanderheijden1999}, Fig.\,\ref{figvdH}b). The dotted black line shows model predictions without the across-channel incoherence interference, so that detection was purely determined by the ERB~=~79\,Hz Gammatone filter centered at 500\,Hz. Despite the large deviations between and within experiments, the model predictions involving the incoherence interference captured the shape of the decreasing thresholds with increasing inner-band bandwidth.

\section{Discussion}	
As long as the masker coherence is fairly constant across frequency bands, experiments on binaural detection can be explained purely on the basis of the coherence $|\gamma|$ defined by a 79\,Hz wide Gammatone filter at ${f_c = 500\,\textrm{Hz}}$ \citep{rabiner1966,encke2021}. This includes fully coherent broadband noise maskers \citep{hirsh1948,vandepar1999}, mixtures of correlated and uncorrelated noise \citep{robinson1963, pollack1959, bernstein2014}, and experiments where the interaural coherence of the masker is reduced by an ITD \citep{langford1964, rabiner1966, bernstein2020c}. However, the on-frequency coherence does not account for thresholds obtained with maskers where these properties change substantially across filter bands. Specifically, the single-channel model version as proposed in \citet{encke2021} is neither able to predict all of the threshold differences between SDN and ODN nor experiments like \citet{marquardt2009} and \citet{kolarik2010} that involve IPD transitions in the masker spectrum (see dashed lines in Figs.\,\ref{figvdH}, \ref{figMc}, \ref{figKol}.

\citet{marquardt2009} hypothesized across-channel processing in the binaural system to explain the reduced binaural benefit under such conditions. Here, we extended the analytical model by \citet{encke2021} to a multi-channel numerical signal-processing model with interference of IPD fluctuations. With across-frequency interference only caused by IPD fluctuations, i.e. by off-frequency incoherence, the proposed model differs from approaches assuming wider binaural filters (e.g. \citealp{vanderheijden1999, kolarik2010}). For stimuli with spectrally constant coherence and masker-target phase relations, like SDN and all conditions simulated by \citet{encke2021}, the incoherence interference has no effect and the model operates on the standard filter bandwidths of its peripheral filterbank. Modeling an interference process, our approach also differs from the symbolic model suggested by \citet{marquardt2009}, which sums interaural cues after binaural interaction and, with their implementation, is also different from wider filters. Their across-frequency processing still causes a stronger damping of binaural sensitivity with increasing masker ITD, which is not seen in the data.

The proposed concept of a detrimental incoherence interference is comparable to modulation detection interference, as shown and discussed by, e.g., \citet{yost1989} and \citet{oxenham2001}. Similar to the proposed across-channel incoherence interference this is modeled by modulation patterns interacting across channels, while energetic spectral masking properties are spectrally limited by peripheral filters \citep{piechowiak2007, dau2013}. Furthermore, a similar process is thought to underlie binaural interference as observed by, e.g., \citet{bernstein1995, best2007,mcfadden1976}.

The dataset of \citet{vanderheijden1999} contains both SDN and ODN, and is therefore the critical challenge for binaural detection models\footnote{
	The most comprehensive simulation of dichotic tone in noise detection thresholds using a cross-correlation-based model is by \citet{bernstein2017a}. It is not expected to simulate the ODN detection thresholds of \citet{vanderheijden1999} with a good accuracy, because an ERB of at least 130\,Hz is necessary. Other ODN stimuli, used experimentally by \citet{bernstein2015}, were included in the model test battery by \citet{bernstein2017a}. Those ODN stimuli, however, differed in several ways from the former. First, the target frequency is 250\,Hz, compared to 500 Hz in \citet{vanderheijden1999} and in all other studies here simulated. Second, instead of fixing the target tone to $S_0$ or $S_{\pi}$, the target is delayed by the same amount as one of the two noises, i.e. $(N_0)_{\pm \textrm{ITD}}(S_{\pi})_{\textrm{ITD}}$. Such an approach is useful for SDN, as it ensures a constant $\pi$ difference between the IPDs of the noise and of the tone. For ODN, however, the IPD of the second noise relative to the tone is offset from $\pi$ by 2xITD. This type of stimulus therefore causes an even more complex ITD-dependence of threshold, which offers no advantage over the ODN from \cite{vanderheijden1999} for filter estimation. With both definitions, corresponding SDN and ODN stimuli can be generated only if the ITD is an integer or a half-integer multiple of the target period (i.e. ITD~=~$n/2f$, $n \in \mathbb{N}$). In \citet{bernstein2015} (their Figure 1, Panel a) these are the two data points at ITD~=~2 and 4\,ms. SDN and ODN thresholds are, however, very similar at those points. Third, the masker bandwidth is 50\,Hz. For such a masker bandwidth smaller than the peripheral filter width, neither \citet{vanderheijden1999} nor our model would predict a considerable threshold difference between SDN and ODN at an ITD of 2 and 4\,ms, since there are no off-frequency regions of considerably lower coherence.}. 
Both \citeauthor{vanderheijden1999}' and our model simulate the data very accurately. Therefore, the discussion focuses on consequences and plausibility of the two different concepts.

The bandwidth of the signals immediately prior to binaural interaction dictates the temporal coherence and thus the decline of BMLD with increasing noise ITD in the absence of internal ITD compensation \citep{langford1964, rabiner1966, vanderheijden1999, dietz2021}. To date, two of the arguably most comprehensive datasets of dichotic tone-in-noise detection, \citet{vanderheijden1999} and \citet{bernstein2020c}, have self-reported mutually exclusive requirements for the binaural bandwidth (ERB~=~130$\ldots$180\,Hz vs. ERB~$\leq$~100\,Hz at 500\,Hz).

A variety of studies aim to estimate the bandwidth at the binaural input stage by means of dichotic tone-in-noise detection, but no consistent picture emerges. There is, for example, a difference in estimated bandwidth between band-widening and notched-noise BMLD data, and between stimuli with different flanking bands (e.g., \citealp{kolarik2010}). Particularly this stimulus-type dependence of the "apparent bandwidth" challenges the assumption that all stimuli are processed by the same system. To us, the most reasonable "unifying" explanation is that filter properties arise from the basilar membrane and also the binaural system can make full use of this spectral resolution. The observation that there is less spectral resolution in some cases is then best explained by an across-frequency process for certain stimulus features -- but in contrast to wider filters it is not affecting all features. The proposed incoherence interference may be this missing across-frequency process. At least it appears to reduce or even eliminate inconsistencies in estimating the bandwidth from various binaural detection experiments.

Another mechanism which has been proposed in the context of bandwidth estimation is an optimal combination of target detectability across frequency channels. Masking patterns in dichotic band-widening experiments have a knee-point at larger bandwidths than their diotic counterparts \citep{vandepar1999, bourbon1965}. \citet{vandepar1999} hypothesized that in narrowband maskers, the similar signal-to-noise ratio (SNR) across frequency channels can be exploited to reduce masking. A model which includes such a mechanism  \citep{breebaart2001b} accounts for the band-widening masking pattern using standard filter bandwidths (i.e. bandwidths as proposed by \citealp{glasberg1990}). It also correctly predicts that the knee-point is only shifted to a higher bandwidth if the masker is fully or almost fully correlated \citep{vanderheijden1998a}. 

Most recent binaural models, such as \citet{bernstein2017a} and \citet{encke2021} already assume a bandwidth as narrow as the peripheral bandwidth. This is also in line with direct measurements of the bandwidth in ITD-sensitive inferior colliculus neurons in cats by \citet{mclaughlin2008}: For delayed noise, as used by \citet{vanderheijden1999}, they found that damping of the cross-correlation function corresponds to the peripheral bandwidth at the respective center frequency.

With the present implementation, the binaural pathway parameters ($\hat{\rho}$, $\sigma_b$, $\sigma_w$) had to be adjusted slightly between conditions with $S_{\pi}$ targets and conditions with $S_0$ targets (see Table \ref{tabexp}). This is due to the binaural system's sensitivity depending on the baseline IPD \citep{hirsh1948}. An angular compression of the decision variable space $\{\zeta\}$ at large IPDs is a possible model extension. Delay-line models can account for this dependence with a corresponding $p(\tau)$ function which defines the sensitivity of the model as a function of its internal delay. However, they then incorrectly predict better unmasking with $N_{\textrm{ITD}} S_0$ compared to $N_{\pi} S_0$ when ITD~=~$T/2$ \citep{breebaart1999}. Simulating the data of \citet{marquardt2009} required slightly different parameter values because their listeners obtained different thresholds compared to \citet{vanderheijden1999} for identical stimuli. This may be due to the different number of presented intervals. Identical model parameters were used for the $S_\pi$ conditions of \citet{vanderheijden1999} and \citet{kolarik2010}. 

\section{Conclusion}
Interaural incoherence interference enables the presented binaural model to simulate  detection thresholds both for maskers with a spectrally constant and with a spectrally modulated coherence.
Employing auditory filters with monaurally estimated bandwidth \cite{glasberg1990}, it predicts the reduced unmasking in opposingly-delayed noises \citep{vanderheijden1999} compared to regular delayed noise.
The concept can help to resolve the inconsistency that binaural models require filter bandwidths as estimated monaurally for most data sets \citep{bernstein2017a, bernstein2020c}, but at least 1.6 times wider filters for broadband opposingly delayed noises \cite{vanderheijden1999} and other spectrally complex maskers \cite{verhey2020}.

The main consequence of using a standard filter bandwidth is that the decline of the binaural benefit with masker ITD can be simulated without internal ITD compensation, as first suggested by \citet{langford1964}.

	\section*{Acknowledgments}
	We thank Steven van de Par for fruitful discussions and comments on the manuscript as well as Matthew J. Goupell and two anonynomous reviewers for their very helpful comments. This work was funded by the Deutsche Forschungsgemeinschaft (DFG, German Research Foundation) – Projektnummer 352015383 – SFB 1330 B4.


\appendix
\section{Derivation of cross-power spectral density in opposingly-delayed noise}
\label{appderiv}
In ODN, two two-channel signals $u(t) = [u(t)\  u(t+ \textrm{ITD})]$ and $z(t) = [z(t)\ z(t-\textrm{ITD})]$ with opposite ITDs, ITD and -ITD, are summed. The cross-power spectral density (CPSD) functions are
\begin{equation}\label{eqcpsd}
	\begin{split}
		& S_{UU}(\omega) = 0.5 e^{i  \,\textrm{ITD}\, \omega},\\
		& S_{ZZ}(\omega) = 0.5 e^{-i  \,\textrm{ITD} \,\omega}.
	\end{split}
\end{equation}
The power spectral density is 0.5 $1/$Hz each, so that the ODN has the same energy as the SDN. Summation of the time signals is equivalent to a summation of their CPSD functions, which leads to 
\begin{equation}\label{eqcos1}
	S_{UZ} = S_{UU}(\omega) + S_{ZZ}(\omega) = \cos(\omega\,\textrm{ITD}).
\end{equation}

This resulting cosine pattern is determined by the sum of the CPSDs' phases adding up or canceling each other at different frequencies. This normalized CPSD $C(\omega)$ represents the coherent energy of the signals as a function of frequency \citep{gardner1992},
\begin{equation}\label{eqcos2}
	C(\omega) = \frac{\left| S_{UZ}(\omega)\right|}{\sqrt{S_{UU}(\omega) S_{ZZ}(\omega)}} = |\cos(\omega \,\textrm{ITD})|.
\end{equation} 
If $|\gamma(\tau)|$ is based on an ensemble average, then $C(\omega) = \mathcal{F}\{|\gamma(\tau)|\}$, with  $\mathcal{F}\{\bullet\}$ the fourier transform.
As a continuous function of $\omega$ it gives a coherence for any frequency $\omega$ representing an infinitesimally small bandwidth, illustrated as continuous lines in Fig.\,\ref{figExplain}d. The coherence for peripherally filtered, i.e. finite-bandwidth signals is an average of the frequencies’ normalized CPSDs $C(\omega)$. The coherence decreases with increasing ITD and increasing bandwidth, as illustrated by the bars in Fig.\,\ref{figExplain}d.
Two superimposed noises with ITD~=~$\pm$2\,ms are in phase at 500\,Hz. At 625\,Hz, however, they have IPDs of  $\pi/2$ and $-\pi/2$, respectively. The  coherence between left and right signals at 625\,Hz is therefore zero.




\end{document}